\documentclass[prd,preprint]{revtex4}

\usepackage{graphicx}
\usepackage{dcolumn}
\usepackage{bm}

\begin{document}
\title{New Experimental Limits on \\ 
Macroscopic Forces Below 100 Microns}
\author{Joshua C. Long$^{*}$, Hilton W. Chan$^{**}$, Allison B. Churnside, Eric A. Gulbis,
  Michael C. M. Varney, and John C. Price}
\affiliation{Department of Physics, University of Colorado, Boulder CO}
\date{September 30, 2002}

\begin{abstract}
Results of an experimental search for new macroscopic
forces with Yukawa range between 5 and 500 microns are presented.  
The experiment uses
1~kHz mechanical oscillators as test masses with a stiff
conducting shield between them to suppress backgrounds.  No signal
is observed above the instrumental thermal noise after 22
hours of integration time.  These results provide the
strongest limits to date between 10 and 100 microns, improve on
previous limits by as much as three orders of
magnitude, and rule out half of the remaining parameter space for
predictions of string--inspired models
with low--energy supersymmetry breaking.  New forces of four times 
gravitational strength or greater are excluded at the 95\% confidence 
level for interaction ranges between 200 and 500 microns.

\noindent $^{*}$ Present address: Los Alamos Neutron Science Center, LANSCE-3,
MS-H855, Los Alamos, NM 87545, USA

\noindent $^{**}$ Present address: Physics Deparment, Stanford University,
Stanford, CA 94305, USA 
\end{abstract}

\maketitle

\section{Introduction}
Experimental tests of Newtonian gravity and searches for new weak
forces in addition to gravity have been conducted over length scales ranging
from light--years down to laboratory distances.  Based on recent
results~\cite{adelberger} and reviews~\cite{fischbach}, new forces 
with strength weaker than or comparable to gravity have been excluded 
over distances ranging between 200~$\mu$m and a light--year.  To date, 
only the single experiment in Ref.~\cite{adelberger} has attained
gravitational sensitivity below 1~mm, and limits on the strength of 
new interactions
increase very rapidly below 100~$\mu$m~\cite{bordag}.  
The sub--millimeter range attracts
continuing experimental interest because many recent theoretical attempts
at the unification of fundamental forces predict specific new
phenomena in this regime.  We present the results of an experiment
which has attained a maximum sensitivity of about four 
times gravitational strength above 200~$\mu$m, and which provides 
the most sensitive limits on new forces between 10 and 100 microns.
\subsection{Current Limits}
Results from experimental searches for new macroscopic forces are most
commonly parameterized by a Yukawa interaction.  The potential due
to gravity and an additional Yukawa--type force between two mass densities
$\rho_{1}(\vec{r}_{1})$ and $\rho_{2}(\vec{r}_{2})$ can be written
\begin{equation}
V = -\int{d\vec{r}_{1}}\int{d\vec{r}_{2}}\frac{G\rho_{1}(\vec{r}_{1})\rho_{2}(\vec{r}_{2})}{r_{12}}[1+\alpha\exp(-r_{12}/\lambda)],  
\end{equation}  
where $G$ is the Gravitational constant, $r_{12}$ is the distance
between $\vec{r}_{1}$ and $\vec{r}_{2}$, $\alpha$ is the strength of the Yukawa interaction
relative to gravity, and $\lambda$ is the range.  Current
experimental limits on $\alpha$ between
1~$\mu$m and 1~cm are shown in Fig.~\ref{fig:limit}, together with
several recent theoretical predictions of new phenomena.  

All previously published limits in Fig.~\ref{fig:limit} (bold, dashed curves) 
are obtained from torsion balance 
experiments.  These curves represent 95\% confidence level limits with
the exception of the result from the Irvine 2--5~cm null
experiment~\cite{hoskins}, which is a 1~$\sigma$ limit.  
The torsion balance
experiment at the University of Washington has attained
gravitational sensitivity or better for ranges above 
200~$\mu$m~\cite{adelberger}.  Below 20~$\mu$m the best previous limit is 
derived from the Casimir force measurement by
Lamoreaux~\cite{lamoreaux}; the curve shown in Fig.~\ref{fig:limit}
is based on the analysis by the
authors~\cite{long,mostepanenko}. From the figure, 
the previous experimental limits allow for 
forces in nature several million times stronger than gravity at ranges
as large as 20~$\mu$m.
\subsection{Motivation}
In addition to the unexplored parameter space, there are many specific
motivations to search for new effects below
100~$\mu$m~\cite{disclaimer}.  Several arise from the 
decades--long effort to describe gravity and the
other fundamental interactions in a unified theoretical framework.
The leading candidates for such a theory are string or M theories,
which must be formulated in more than three spatial dimensions.  In
the traditional view, the extra dimensions are taken to be compactified 
on a scale on the order of the Planck length ($L_{P} \approx
10^{-35}$~m or $L_{P}^{-1} \approx 10^{19}$~GeV), a fundamental 
scale at which the unification is expected to occur.  Recently, a 
class of models has been discovered in which the unification can occur
at much lower energies, near the weak scale 
($L_{W}^{-1} \approx 1$~TeV)~\cite{hewett}.  In this picture, perhaps the most 
frequently cited development in particle physics in recent years, the 
discrepancy between the Planck scale and the weak scale 
(the so--called hierarchy problem) is removed as a consequence of 
some of the extra dimensions remaining large and accessible only 
to gravity, with the Standard
Model fields confined to the usual three
dimensions~\cite{arkani}.  The size $R$ of the extra dimensions is given by
$R^{n}=L_{W}^{2+n}/L_{P}^{2}$, where $n$ is
the number of compact dimensions.  For $n = 1$, $R \approx 10^{13}$~m,
clearly ruled out by astrophysics.  However, the choice of $n = 2$ 
implies $R \approx 1$~mm, with the consequence that the gravitational
potential will behave according to a $1/r^{3}$ law below this scale.  As
the scale $R$ is approached from above, Yukawa corrections are
predicted; the model illustrated in
Fig.~\ref{fig:limit} predicts $\alpha = 4$~\cite{floratos}.     

Another class of predictions arises in superstring theories in which
supersymmetry (SUSY) is broken at low energies.  These models are of
interest as unification scenarios that account for the absence of 
flavor--changing neutral currents.
Superstring theories generally contain gravitationally--coupled scalar fields
called moduli, which are massless at the string scale but which can
acquire mass from the same process which breaks SUSY.  For models in
which SUSY is broken by a gauge--mediated process between 10 and 100~TeV, the moduli acquire masses at the sub-eV level corresponding to interaction
ranges between 100~$\mu$m and 10~cm, with values of $\alpha$
as high as $10^{6}$~\cite{dimopoulos}.  Predictions for the largest
such effects (for the moduli which couple to the strange quark and
gluon) with the uncertainties are shown in Fig.~\ref{fig:limit}. In a
related model (radius modulus) in which SUSY is broken near 1~TeV 
via weak--scale compactification~\cite{antoniadis}, the moduli can
acquire Compton 
wavelengths in the range from 10~$\mu$m to 1~mm.  The associated 
strength is predicted to 
be $\alpha = 1/3$, as shown in Fig.~\ref{fig:limit}.

The dilaton is another scalar field predicted by string theories,
which can also acquire mass from SUSY breaking.  Due to its more
universal couplings, it is expected to acquire a mass too large to be
observable in the scenario in Ref.~\cite{dimopoulos}, but other authors
take the dilaton mass to be an unknown parameter.  The estimated
coupling for the dilaton from Ref.~\cite{ellis} is shown in
Fig.~\ref{fig:limit}, where previous experimental limits exclude a
mass less than about $3.2 \times 10^{-3}$~eV; more recent calculations 
can be found in Ref.~\cite{kaplan}.      

The axion, a light pseudoscalar boson motivated by the strong CP
problem of the Standard Model, is also predicted to mediate
macroscopic forces.  Astrophysical and laboratory bounds have left an
allowed window for the axion mass corresponding to 
200~$\mu$m $< \lambda <$ 20~cm~\cite{rosenberg}.  The coupling of the 
associated long--range force
between unpolarized test masses is constrained by measurements of the 
neutron electric dipole moment to the region indicated in
Fig.~\ref{fig:limit}.  For pseudoscalars which are not derivative
coupled (unlike the axion), double exchange leads to a $1/r^{3}$
potential between unpolarized test masses; short range experiments may
set better limits on pseudoscalars via this effect~\cite{krause}.    

Additional predictions are motivated by the cosmological constant
($\Lambda$) problem, the discrepancy between the observed flatness 
of the universe versus the extreme curvature expected from the 
vacuum energy contributions of the Standard Model fields.  Recent
models~\cite{sundrum,beane} assert that if a sufficient fraction of the
energy density in the universe is in the form of vacuum energy,
consistency with local field theory implies the existence of new interacting
quanta with mass on the order of $\Lambda^{1/4}$.  The predictions in 
Fig.~\ref{fig:limit} are derived from the conservative
assumption of cosmological vacuum energy density $\rho_{\Lambda}$ 
in the range $0.1\rho_{c}<\rho_{\Lambda}<\rho_{c}$, where $\rho_{c}$ is
the critical density of the universe.
\section{Experiment}
\subsection{Apparatus}
\begin{figure}
\includegraphics[width=10cm]{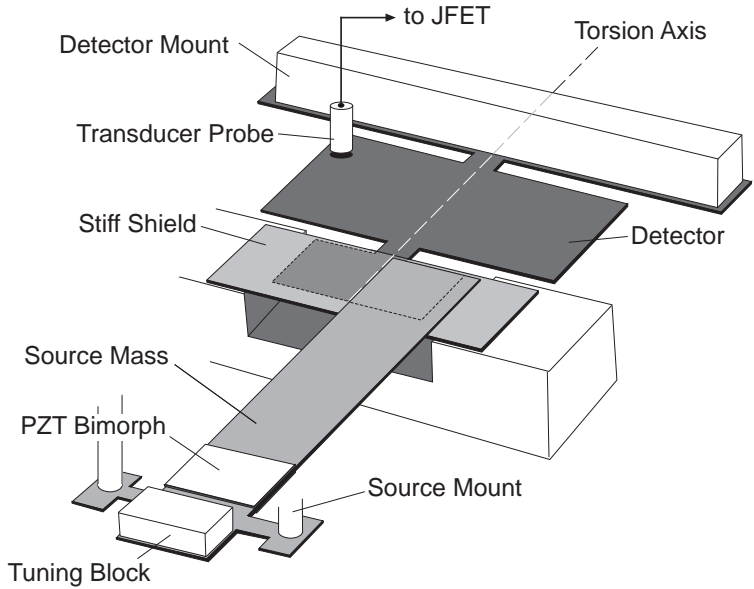}
\caption{\label{fig:expt} Central components of the apparatus}
\end{figure}

The present experiment is illustrated in Fig.~\ref{fig:expt}.  The
test masses consist of a planar resonant detector mass and a planar
source mass driven at the detector resonant frequency.  
This geometry is chosen 
to concentrate as much test mass density
as possible at the length scale of interest.  It is also largely
null with respect to $1/r^{2}$ forces and so is effective in
suppressing the Newtonian background in the context of a new force
search.  The source mass consists
of a $35~\mbox{mm} \times 7~\mbox{mm} \times 0.305~\mbox{mm}$
node--mounted tungsten reed.  It is driven at its second cantilever
mode by a PZT
bimorph at a frequency carefully matched to the 1~kHz 
resonant frequency of a normal mode of the detector mass.  
The detector mass consists of a 0.195 mm thick tungsten torsional 
oscillator of a double--rectangle design.  The small rectangle measures 
$11.455~\mbox{mm} \times 5.080~\mbox{mm}$.  For the resonant mode of
interest (the 5th normal mode) the two rectangular sections 
counter--rotate about the torsional axis, with most
of the amplitude confined to the motion of the small rectangle.  This
double--torsional arrangement provides significant isolation
of the small rectangle from the detector mount, reducing
mode damping~\cite{kleiman}.
When data are collected, the front edge of the source mass is 
aligned with the back edge of the small detector rectangle, and a 
long edge of the source is aligned above the detector torsion axis.  
This geometry (called ``fiducial'' below) maximizes the on--resonance 
torque on the detector if a
mass--coupled force is present between the source and detector.

A stiff conducting shield suspended between the test masses 
suppresses electrostatic and acoustic backgrounds.  The shield
consists of a 0.060~mm thick sapphire plate with a 100~nm gold
plating.  The shield is mounted on a brass support and secured on
opposite ends with epoxy.

The test masses and shield support are each suspended from modified 
optical mirror mounts, which are used as a tilt stages to 
level each element in the horizontal plane.  
Each tilt stage is in turn mounted to the bottom segment of a 
vibration isolation stack.  Each stack consists of five solid brass 
disks connected by fine steel
wires, as described in detail elsewhere~\cite{chan}.  At the 1~kHz
operational frequency of the experiment, each stack provides
an attenuation of approximately 200~dB.  The vibration isolation
stacks of the source and detector masses are suspended from inverted 
3-axis micrometer stages, which provide full translation control.  The
relative positions and coplanarity of the test masses and shield are 
ascertained by
touching the elements against each other, or against a series of
0.5~mm diameter sapphire hemispheres attached to small rods at the end
of the shield support.

To further reduce acoustic backgrounds, the apparatus is placed in a 75 liter vacuum bell jar, which is
pumped to a pressure of approximately $2 \times 10^{-7}$~torr with a liquid
nitrogen--trapped diffusion pump.  In order to isolate the
(non-magnetic) central apparatus from stray fields generated
in the steel pump components, the bell jar is connected to the
diffusion pump via a 1~m long, 10~cm diameter aluminum riser.  The
position--control micrometer stages are manually operated with torque
rods which exit the bell jar via rotary feedthroughs.

In the absence of electromagnetic, acoustic, and vibrational
backgrounds, the experiment is limited by thermal noise due to 
dissipation in the detector mass.  To reduce
this dissipation, the detector mass is annealed at $1300~^{\circ}$C for several
hours in an induction furnace under a helium atmosphere before
installation in the bell jar.  This is observed to increase the
detector mechanical quality factor ($Q$) typically by factors of 5 or more.   

The detector temperature is stabilized with an electronic 
temperature controller using a silicon diode sensor and 
resistive heating element mounted to
the bottom stage of the detector vibration isolation stack.  The
temperature is maintained at 305~K with fluctuations of about 0.1~K.
\subsection{Readout}
\begin{figure}
\includegraphics[width=13cm]{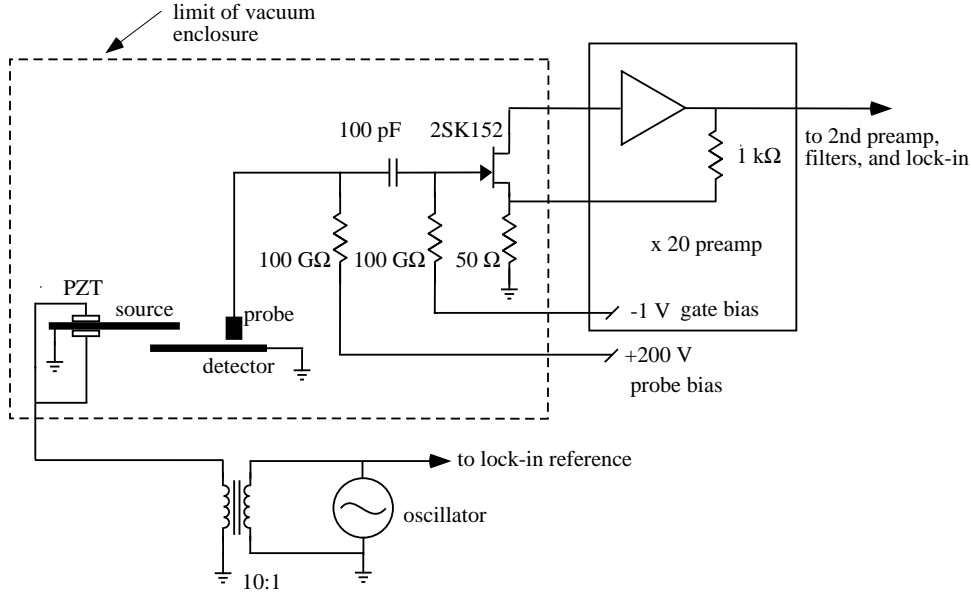}
 \caption{\label{fig:readout} Major components of the detector readout
   and PZT drive}
\end{figure}

The detector readout electronics and source mass PZT drive are illustrated in 
Fig.~\ref{fig:readout}.  Oscillations of the detector mass are read
out with a capacitive
transducer.  The capacitive transducer probe consists of a
2.5~mm diameter brass cylinder supported with its flat end
approximately 0.1~mm above a rear corner of the large rectangle of
the detector mass.  The probe is biased at 200~V through a 
100~G$\Omega$ resistor, whose value must be large to reduce current noise.  
The front end of the 
preamplifier consists of an SK152 JFET located in a small box
immediately above the detector oscillator.  The 100~mK noise
temperature of the SK152 preamplifier ensures that this circuit is more than
sufficient for detecting the 300~K thermal oscillations (amplitude $\approx$
100~fm) of the detector mass.  The JFET preamplifier is followed by a 
second preamplifier (Stanford Research
SR560), filters, and finally a two phase lock--in amplifier. The total
voltage gain from the capacitive probe to the lock--in input is about 1600.
A crystal--controlled oscillator provides a reference signal for the
lock--in amplifier and drives the source mass PZT through a 1:10
step--up transformer.
\section{Data Sample}
\subsection{Force Measurement Data}
The data were recorded in 108 ``cycles'' over five days.
Each cycle consists of 7 runs containing both force measurement data
and diagnostic data.  The diagnostic data are used to continually
monitor the performance of the instrument. For all runs the bandwidth
of the lock-in amplifier was set to 250~mHz (roughly 5 times the
width of the detector resonance) to include the noise power 
of the detector thermal oscillations, which is used for
calibration.  A data sampling rate of 1~Hz was used, corresponding to 
an over-sampling of about a factor of 10.  

Within each cycle, the runs were taken in the following
order:  First, a stable dc bias (usually 5--10~V) was applied 
to the shield to induce a large resonant electrostatic
signal.  A run of 120 samples was taken with the PZT
drive frequency set 30~mHz below the approximate value of the detector
resonant frequency.  Four subsequent runs were taken, each with the PZT
drive frequency incremented by 15~mHz relative to the previous run in order 
to display the detector resonance.  The precise value of the
detector resonant frequency was determined from the five biased runs,
and was then used as the drive frequency for the next run of the cycle.
The shield was then grounded and a run of 720 samples
(12 minutes) was recorded.  This was followed by a 
shorter run of 288 samples (4.8 minutes) taken with the PZT drive set
to 1171.000~Hz (at least 2~Hz below the detector resonance), in order
to monitor the system for non--resonant systematic offsets.  The entire cycle
was repeated indefinitely until
an intervention to service the experiment was necessary.
A total of 108 such cycles were acquired yielding a total of 
77760 on--resonance samples.

\begin{figure}[h]
\includegraphics[width=13cm]{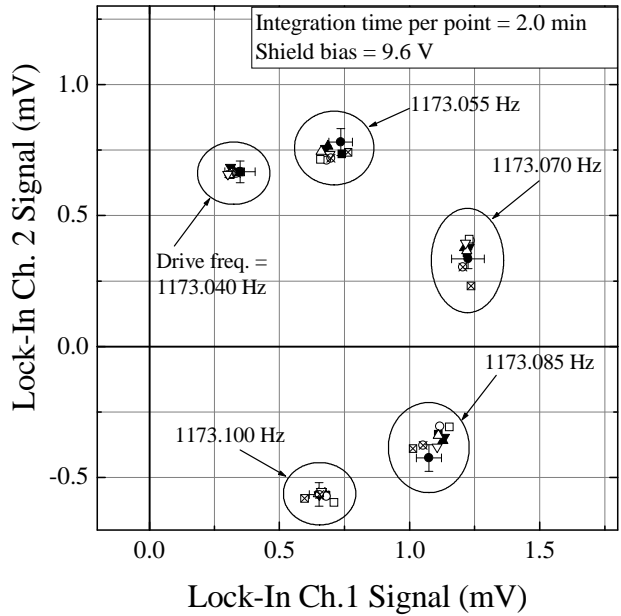}
\caption{\label{fig:diagnostic} Data from the biased runs of 10
  consecutive cycles of the data sample.  Points from separate cycles are
  labeled with
  unique symbols. 1~$\sigma$ error bars are shown for one cycle. Groups of
  points corresponding to a particular drive frequency are circled.}
\end{figure}

A plot of the data from the biased diagnostic runs is shown in
Fig.~\ref{fig:diagnostic} (only data from 10 consecutive cycles are shown for
clarity).  The signals from the two phase--quadrature channels of the
lock-in amplifier are plotted against each other to
show the phase behavior of the signal with drive frequency.  In
all cycles, the signal maximizes at very nearly the same phase and
magnitude, indicating good stability of the detector resonant
frequency, source mass amplitude, and system gain.

\begin{figure}[h]
\includegraphics[width=13.5cm]{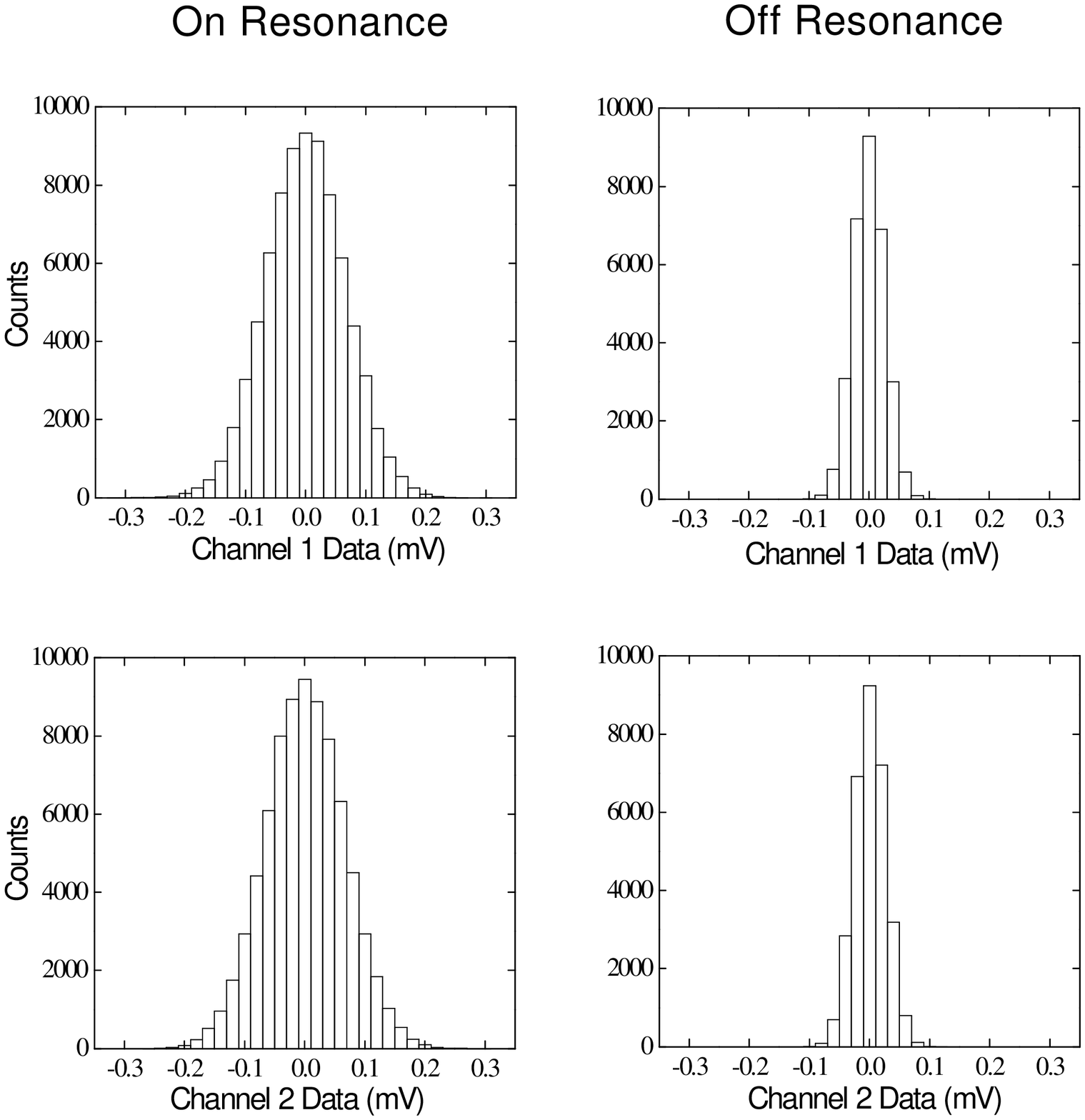}
\caption{\label{fig:raw} Distributions of data samples without shield
  bias. Left--hand plots: Drive (and lock-in
  reference) tuned to detector resonance.  Right--hand plots: Drive
  tuned 2~Hz below detector resonance.}
\end{figure}

\begin{figure}
\includegraphics[width=11cm]{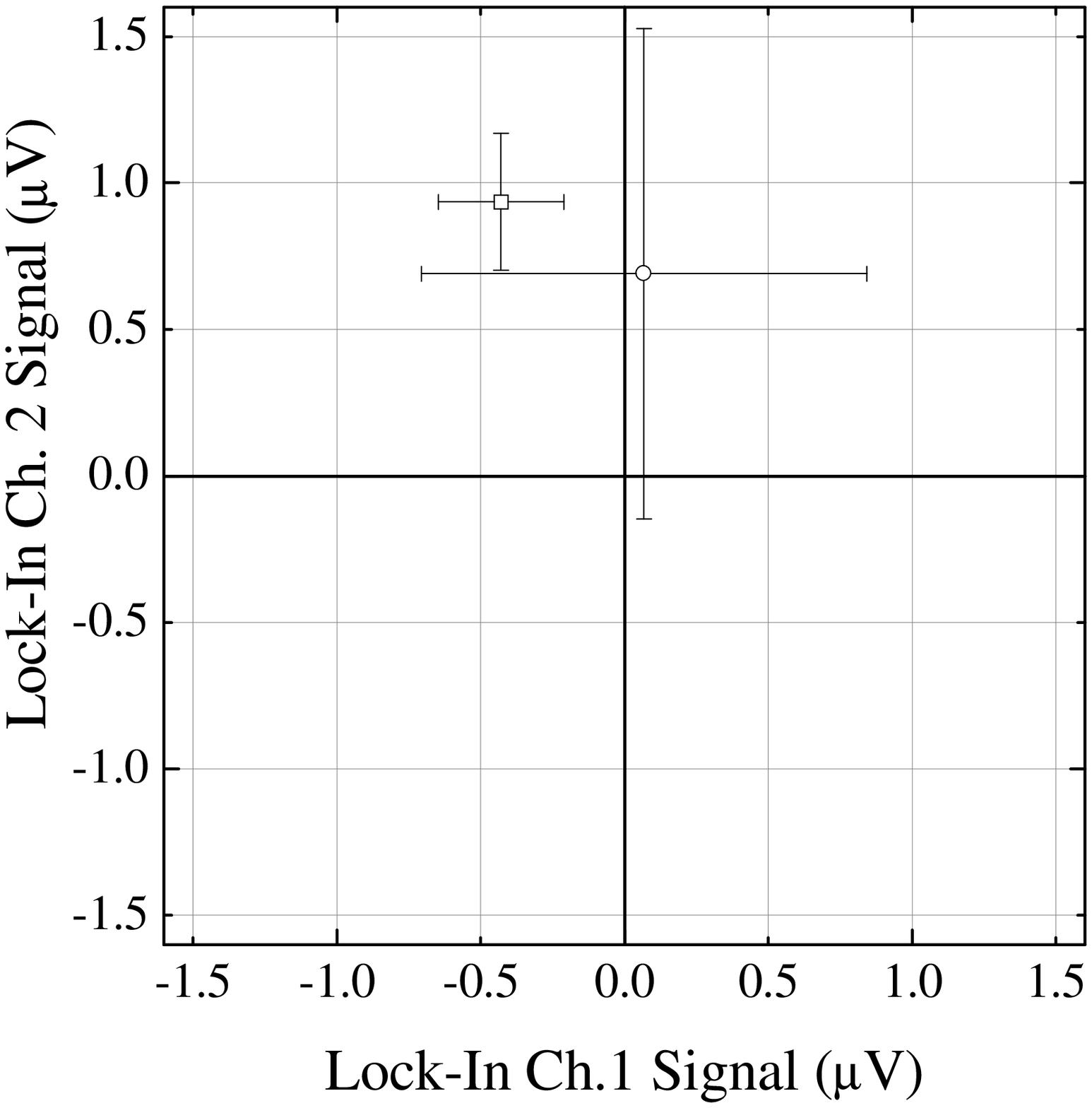}
\caption{\label{fig:summary} Means of the distributions in Fig.~\ref{fig:raw}. 
  with channels plotted against each other to show phase. Error bars are
  1~$\sigma$ standard deviations of the means. The point with the
  larger error bars is the on-resonance mean.} 
\end{figure}

Fig.~\ref{fig:raw} shows histograms of the on-- and off--resonance
unbiased data.  The plots combine data from all 108 on--resonance
and off--resonance runs, each plot displaying data from a 
single channel of the lock-in amplifier.  The data exhibit smooth 
gaussian behavior centered about common means, as expected in 
the exclusive presence of detector thermal noise and amplifier noise.  
The widths of the distributions of the on--resonance data 
(left-hand plots in Fig.~\ref{fig:raw}) are roughly
twice those of the off--resonance data (right-hand plots) due to
the contribution of detector thermal noise.
The means of the distributions in Fig.~\ref{fig:raw} are shown in 
Fig.~\ref{fig:summary}, in which the
data from the separate lock-in amplifier channels are plotted against
each other as in Fig.~\ref{fig:diagnostic}.  The on--and
off--resonance means agree within the 1~$\sigma$ standard deviations
shown, indicating the absence of any resonant force signal.
The offset from the origin of the mean of the off--resonance
distribution is due to leakage of the reference signal internal to
the lock--in.
\subsection{Consistency Checks}
After the acquisition of the session data, several checks are made to 
verify the stability of the results and their consistency with 
known backgrounds.  First, several additional cycles are acquired 
with variations in the test mass geometry, including larger vertical
gaps and different overlap configurations.  No resonant signal is observed 
in any of these sessions.  This strongly disfavors the possibility 
that the observed null
result is due to a fortuitous cancellation of surface potential,
magnetic, and/or acoustic effects, all of which are expected to have
different dependencies on the geometry.

Next, several on--resonance runs are acquired with different
transducer probe 
bias voltage settings and with the source mass drive turned off.
The observed linear dependence of the rms fluctuation in these data 
on the probe bias voltage is consistent with detector motion due
only to thermal noise and rules out additional motion from
transducer back-action noise. This check is
important because the magnitude of the detector thermal motion is used
for calibration.

The data from the diagnostic runs with shield bias voltage applied can
be used to estimate the minimum size of the residual potential
difference between the shield and the (grounded) test masses needed to
produce a resonant signal.  From Fig.~\ref{fig:diagnostic}, the maximum signal
with 9.6~V shield bias is about 1.3~mV.  Further measurements show
that this signal scales as the fourth power of the shield bias voltage, 
as expected for an electrostatic force between source and detector 
mediated by a deflected shield.  Scaling the ratio of the on--resonance
error bar in Fig.~\ref{fig:summary} to the 1.3~mV diagnostic implies
that a residual shield bias of at least 1.5~V would be 
needed to generate a systematic effect above detector thermal noise.
This is about an order of magnitude larger than the measured residual 
potential difference between the shield and test masses.

Magnetic effects can generate background signals though several
mechanisms. The most important effect involves
generation of eddy currents when the source mass moves in an external
magnetic field.  Fields produced by the source eddy currents create eddy
currents in the detector, which then interact with the applied field.
Studies of this effect with
large applied fields show that the induced force varies as the square
of the applied field (as expected), and extrapolation to the ambient
field actually present indicates that this effect should be about 5 times
smaller than the thermal noise limited sensitivity.
\subsection{Auxiliary Measurements}
Before acquiring the session data samples, several measurements are
made to support the analysis of the data for evidence of a net
force.  These include the relative
phases of the motion of the test masses, the detector mechanical quality
factor, and a precise survey of the test mass geometry. 

The Yukawa interaction for the test mass geometry of the
experiment is used to model any potential signal.  Such an interaction
can result in either a purely attractive or repulsive force
between the source and detector masses.  In order to ascertain the phase
of the signal for such a force, the source and detector are placed in
their fiducial positions with an additional vertical offset so that both are
situated above the electrostatic shield.  An electrostatic bias of
1.5~V is applied to the detector mass with a dry cell in order to
induce a large, stable, purely attractive electrostatic coupling between 
the test masses.  A set of runs is taken with small increments of
the PZT drive frequency between successive runs.  The position of the
data point in the complex plane corresponding to the maximum signal as
a function of drive frequency is recorded.  The phase of this signal 
corresponds to that of a purely attractive force.  The phase of a
purely attractive signal is
$189^{\circ}$, just below the negative horizontal axis in the plots in 
Figs.~\ref{fig:diagnostic} and~\ref{fig:summary}.

The detector mechanical quality factor $Q$ is measured by applying a 100~mV
resonant ac signal directly to the detector, switching the detector to
ground, and observing the ring-down signal on the lock-in amplifier
with the reference frequency tuned 1~Hz off the detector resonance.  A
value for $Q$ is obtained from a least-squares fit to the ring-down
waveform.  Several such measurements are made before and after the
data cycles reported here, yielding a $Q$ of $25522\pm 29$.

In the initial step of the test mass geometry survey, the source and 
detector masses are positioned over the electrostatic shield.  The
source is tilted on its horizontal axes to ensure that the front left corner is the
lowest point on the bottom surface.  This corner is then used as a
probe to make a map of the top surface of the detector mass, using a
square grid of points with a point density of 1.0~mm$^{-2}$.  The tilt
stage of the detector mass is then adjusted to optimize the level of
the detector, and the surface is mapped again.  A similar procedure is
then carried out for the electrostatic shield.  Finally, a map of the
bottom surface of the source mass is obtained by touching this surface
off against the sapphire hemisphere probe fixed to the end of the shield
mount.  The level is optimized with the source mass tilt stage, and the 
map is repeated. 

In order to obtain the modeshape and amplitude of the driven source 
mass, the PZT is driven at the resonance of the detector and another 
grid of points is obtained by touching the bottom surface of the source mass
against the sapphire probe.  These measurements are done at
atmospheric pressure and the touch--offs are determined acoustically. 

The system is then closed and brought down to diffusion pump pressure.  The
relative horizontal positions of the source and detector masses are determined
by touching an edge of the source against an edge of the
detector, and then the horizontal positions are set to the fiducial
locations.  Next, the source is lowered to bring the test masses into
momentary contact, and their vertical and horizontal positions are recorded.
The detector mass is then positioned under the electrostatic
shield.  First it is brought as far forward under the shield as
possible (without the detector and shield stacks making physical
contact), then centered between the sides of the shield mount as
determined by touch-off signals.  The same net horizontal translations
required to position the detector are applied to the source,
bringing the test masses into their fiducial positions. 

Finally, the minimum vertical gap is established.  The top surface of the 
detector is brought into momentary contact with the bottom surface of
the shield, then backed off by 20~$\mu$m.  The bottom surface of the source is
brought into momentary contact with the top surface of the shield, and
the source is backed off by 20~$\mu$m (the amplitude at the end of the
source mass) plus an additional 10~$\mu$m for safety. 
\subsection{Calibration}
Conversion of the voltage data to observed force is achieved by
direct comparison of the mean signal to thermal noise.
Comparison of the root mean square thermal oscillations of the detector
mass to the mean displacement of the detector when driven by the
source mass leads to the expression
\begin{equation}
\overline{V^{D}} = \sqrt{\overline{\left|V^{T}\right|^2}} 
  \frac{Q}{\omega_{0}\sqrt{k_{B}T\rho_{d}}}\frac{\int d^{3}\vec{r}^{\prime}\vec{z}^{F}(\vec{r}^{\prime})\cdot
  \vec{f}(\vec{r}^{\prime})}{\sqrt{\int d^{3}\vec{r}^{\prime}\left|\vec{z}^{F}(\vec{r}^{\prime})\right|^2}},
\label{eq:measuredvolt}
\end{equation}
where $\overline{V^{D}}$ is the mean of the distribution of voltages
in the data sample, $\sqrt{\overline{\left|V^{T}\right|^2}}$ is the 
component of the standard deviation of the same data sample due to 
detector thermal noise, $k_{B}$ is Boltzmann's constant, and 
$Q, \omega_{0}, T$, and $\rho_{d}$ are the
detector mechanical quality factor, resonant frequency, temperature,
and density.  Both $\overline{V^{D}}$ and 
$\sqrt{\overline{\left|V^{T}\right|^2}}$ are evaluated at a phase of
$189^{\circ}$, corresponding to a purely attractive force.  
The expression $\vec{z}^{F}(\vec{r}^{\prime})\cdot
\vec{f}(\vec{r}^{\prime})$ is the
projection of the driving force density $\vec{f}(\vec{r}^{\prime})$
times the free modal
displacement $\vec{z}^{F}(\vec{r}^{\prime})$ of the detector at 
arbitrary point $\vec{r}^{\prime}$ on the detector, and 
$\left|\vec{z}^{F}(\vec{r}^{\prime})\right|^2$ the mean square 
free modal displacement; both are integrated over the detector
volume.  A derivation of Eq.~\ref{eq:measuredvolt} is given 
in Appendix A.  This expression can be used to convert 
$\overline{V^{D}}$ directly to
observed force.  Alternately, the instrument may be calibrated using a
calculable electrostatic force or by using the reciprocity of the
transducer.  Both methods agree with calibration based on the
thermal motion, but the uncertainties involved are substantially larger.

\section{Analysis}
Generally, the constraints on the Yukawa parameter $\alpha$ for a given 
range $\lambda$ are obtained by ascertaining how large of
a Yukawa interaction between the test masses could be present and
still be consistent with the data in Fig.~\ref{fig:summary}.  We
present first a simplified analysis based on an idealized geometry and
neglecting all systematic effects.  As the
data are consistent with a null result limited by detector thermal
noise, this preliminary estimate can
made by comparing the hypothetical Yukawa force between the source and
detector with the detector thermal noise.

Following the previous analysis of this experiment in
Ref.~\cite{long}, the detector modeshape is taken to be a pure rotation
and the source mass motion a pure translation normal to its surface.
The source mass is driven at the detector resonant frequency, and only
the Fourier amplitude of the Yukawa torque at this frequency is
effective in driving the detector.  From
Ref.~\cite{long}, the torque amplitude is
\begin{equation}
|N_{Y}(\omega_{0})|=2\pi\alpha G\rho_{d}\rho_{s}A_{d}R\lambda^{2}
I_{1}(dz_{s}/\lambda)\exp{(-g_{sd}/\lambda)}
[1-\exp{(-t_{d}/\lambda)}][1-\exp{(-t_{s}/\lambda)}].
\label{eq:yukawa}
\end{equation}
Here, $\rho_{s}$ is the source mass density, $A_{d}$ is the area of
the detector under the source
mass (half the area of the small rectangle), $R$ is the distance from 
the edge of the detector to the
torsion axis (equal to $w_{d}/2$; see Table~\ref{tab:systematics2} 
below), $dz_{s}$ is the source mass amplitude,
$g_{sd}$ is the average source--detector gap,  
$I_{1}(dz_{s}/\lambda)$ is the modified Bessel function, and $t_{d}$ 
and $t_{s}$ are the detector and source thicknesses.  This expression 
neglects edge effects and does not include the true cantilever
modeshape of the source mass.

The thermal noise torque in the experimental bandwidth on the small
detector rectangle is
found from the mechanical Nyquist expression:
\begin{equation}
N_{T}=\sqrt{\frac{4k_{B}T}{\tau}\left(\frac{mR^{2}\omega_{0}}{3Q}\right)}.
\label{eq:noise}
\end{equation}
Here, $m$ is the mass of the small rectangle, $\tau$ is the
inverse of the bandwidth, and $Q$ is the detector quality factor.

The signal--to--noise ratio is the ratio of Eq.~\ref{eq:yukawa}
to Eq.~\ref{eq:noise}.  Setting this equal to unity and solving
for $\alpha$ yields an approximation to the experimental
limit curve shown in Fig.~\ref{fig:limit}.  Evaluating this expression using
the mean values of the parameters listed in
Tables~\ref{tab:systematics} and~\ref{tab:systematics2} below, assuming an
integration time of 77760~s and an average source amplitude equal to
1/2 the measured tip amplitude, yields a limit
in the ($\alpha, \lambda$) space stronger than the result in Fig.~\ref{fig:limit} by a small numerical factor for
ranges $\lambda > 50~\mu$m.  The discrepancy increases to
about two orders of magnitude at $\lambda = 5~\mu$m due
to the high sensitivity of the Yukawa exponential to systematic
errors in the source amplitude and the average gap.

To take the precise geometry and the systematic error into account, the
Yukawa force between the test masses is computed numerically and
constraints on the Yukawa strength $\alpha$ in Fig.~\ref{fig:limit}
are calculated using a maximum likelihood technique~\cite{hagiwara}.  
In this approach, the 
interval $[\alpha_{\mbox{\scriptsize
    lo}},\alpha_{\mbox{\scriptsize up}}]$ which
contains the true value of $\alpha$ with probability $CL$ is given by
\begin{equation}
CL = \int^{\alpha_{\mbox{\scriptsize up}}}_{\alpha_{\mbox{\scriptsize
      lo}}}p(\alpha|\mbox{\boldmath$x$})d\alpha,
\label{eq:CL}
\end{equation}
where $p(\alpha|\mbox{\boldmath$x$})$ is the posterior probability
density function (p.d.f.) for $\alpha$ given the experimental 
data {\boldmath$x$}. The posterior p.d.f. is in turn calculated 
from Bayes' theorem,
\begin{equation}
p(\alpha|\mbox{\boldmath$x$}) = \frac{L(\mbox{\boldmath$x$}|\alpha)\pi(\alpha)}{\int^{\infty}_{-\infty}L(\mbox{\boldmath$x$}|\alpha^{\prime})\pi(\alpha^{\prime})d\alpha^{\prime}},
\label{eq:pdf}
\end{equation}
where $L(\mbox{\boldmath$x$}|\alpha)$ is the likelihood function 
and $\pi(\alpha)$ is the prior p.d.f. for $\alpha$.

In order to account for the effects of various systematics such as
test mass geometry, density, and other mechanical properties, 
the likelihood function in Eq.~\ref{eq:pdf} is replaced by the
expression 
\begin{equation}
L^{\prime}(\mbox{\boldmath$x$}|\alpha) = \int
L(\mbox{\boldmath$x$}|\alpha,\mbox{\boldmath$\nu$})\pi(\mbox{\boldmath$\nu$})d\mbox{\boldmath$\nu$},
\label{eq:likelihood}
\end{equation}
where {\boldmath$\nu$} represents the set of systematic variables and
$\pi(\mbox{\boldmath$\nu$})$ is their prior p.d.f..
In Eq.~\ref{eq:likelihood}, 
$L(\mbox{\boldmath$x$}|\alpha,\mbox{\boldmath$\nu$})$ is given by
\begin{equation}
L(\mbox{\boldmath$x$}|\alpha,\mbox{\boldmath$\nu$})
= \left(\frac{1}{\sigma
    \sqrt{2\pi e}} \right)^{N} \exp\left[-\frac{(\bar{x}-\mu(\alpha,\mbox{\boldmath$\nu$}))^2}{2(\sigma/\sqrt{N})^{2}}\right],
\label{eq:prod}
\end{equation}
where $\bar{x}$ is the average of the voltages {\boldmath$x$} in the data
session, $\sigma$ is their standard deviation, and $N$ is the effective
number of uncorrelated samples.  The term
$\mu(\alpha,\mbox{\boldmath$\nu$})$ is the {\it predicted} mean 
voltage for a given $\alpha$ and set of systematics {\boldmath $\nu$}.
It is equivalent to $\overline{V^{D}}$ in Eq.~\ref{eq:measuredvolt}, where
$\vec{f}(\vec{r}^{\prime})$ in that equation is understood to be 
the {\it theoretical} value of the Yukawa force density for the experimental
test mass geometry.  A more detailed expression for the
likelihood function is derived in Appendix B.
\subsection{Evaluation of the Likelihood Function}
Eq.~\ref{eq:likelihood} is evaluated by Monte Carlo integration over
the systematics {\boldmath $\nu$}.  For a fixed value of the Yukawa range
$\lambda$, 400 points are thrown in the sample space
defined by the volume $d\mbox{\boldmath$\nu$}$; the integrand is calculated
for each point and added to a running total.  

The parameter $\bar{x}$ in Eq.~\ref{eq:prod} represents the mean of the 
on--resonance data with respect to the mean of 
the off--resonance data (which measures the effective zero), at the
phase corresponding to the predicted signal.  To find
$\bar{x}$, all data are first projected onto the phase of the purely
attractive signal. The
mean of the off--resonance data is then subtracted from that of the
on--resonance data.  The value of $\bar{x}$ obtained is $-0.44~\mu$V.

The quantity $\sigma/\sqrt{N}$, equal to the
standard deviation of the mean of the data, is calculated in the following way:
First, the projected data sets are partitioned
into $N_{\mbox{\scriptsize part}}$ sets of equal samples, with
$N_{\mbox{\scriptsize part}}$ large enough so that there is very
little correlation between partitions. The data in
each of the partitions
are averaged separately and the standard deviation of the resulting 
averages $\sigma_{\mbox{\scriptsize av}}$ is computed.  The standard
deviation of the mean is then calculated from 
$\sigma_{\mbox{\scriptsize av}}/\sqrt{N_{\mbox{\scriptsize part}}}$.  
The effective number of uncorrelated samples is the value of $N$
required to make $\sigma/\sqrt{N} =
\sigma_{\mbox{\scriptsize av}}/\sqrt{N_{\mbox{\scriptsize part}}}$.
Finally, values of $\sigma/\sqrt{N}$ obtained 
separately for the on-- and off--resonance data sets are added in 
quadrature.  The resulting value of
$\sigma/\sqrt{N}$ is $0.82~\mu$V.

A list the parameters {\boldmath $\nu$} needed to evaluate
Eq.~\ref{eq:measuredvolt} is given in Table~\ref{tab:systematics}.  
Table~\ref{tab:systematics2} lists additional systematics which specify the
test mass geometry used in the computation of the integral $\int
d^{3}\vec{r}^{\prime}\vec{z}^{F}(\vec{r}^{\prime})\cdot
\vec{f}(\vec{r}^{\prime})$ in Eq.~\ref{eq:measuredvolt}.
\begin{table}[hb]
\caption{\label{tab:systematics}Systematics in
  Eq.~\ref{eq:measuredvolt} for evaluation of likelihood function}
\begin{ruledtabular}
\begin{tabular}{lccr}
Parameter&Mean&Error&Units\\
\hline
Gravitational constant, $G$&$6.673\times
10^{-11}$&$1.0\times10^{-13}$&$\mbox{m}^{3}\mbox{kg}^{-1}\mbox{s}^{-2}$  \\
Boltzmann constant, $k_{B}$&$1.3806503\times 10^{-23}$&$2.4\times 10^{-29}$&J~$\mbox{K}^{-1}$\\
Detector density (tungsten), $\rho_{d}$&$1.93\times 10^{4}$&$1.9\times
10^{3}$&$\mbox{kg}~\mbox{m}^{-3}$\\
Source density (tungsten), $\rho_{s}$&$1.93\times 10^{4}$&$1.9\times 10^{3}$&$\mbox{kg}~\mbox{m}^{-3}$\\
Thermal noise voltage,
$\sqrt{\overline{\left|V^{T}\right|^2}}$&$6.09\times
10^{-5}$&$2.3\times 10^{-6}$&V\\
Mechanical quality factor, $Q$&$2.5522\times 10^{4}$&29&(NA)\\
Resonant frequency, $\omega_{0}/2\pi$&1173.085&0.015&Hz\\
Temperature, $T$&305.0&0.1&K\\
Integrated rms free modeshape, $\sqrt{\int
d^{3}\vec{r}^{\prime}\left|\vec{z}^{F}(\vec{r}^{\prime})\right|^2}$&$5.87\times
10^{-11}$&$5.9\times 10^{-12}$&$\mbox{m}^{5/2}$\\
\end{tabular}
\end{ruledtabular}
\end{table}

\begin{table}
\caption{\label{tab:systematics2}Systematics for evaluation of $\int
d^{3}\vec{r}^{\prime}\vec{z}^{F}(\vec{r}^{\prime})\cdot
\vec{\mbox{f}}(\vec{r}^{\prime})$.  All units are meters.  See text
for definition of parameters.}
\begin{ruledtabular}
\begin{tabular}{lcr}
Parameter&Mean&Error\\
\hline
Detector length, $l_{d}$&$5.0800\times 10^{-3}$&$6.4\times10^{-6}$\\
Detector width, $w_{d}$&$1.14550\times 10^{-2}$&$6.4\times10^{-6}$\\
Detector thickness, $t_{d}$&$1.950\times 10^{-4}$&$6.4\times10^{-6}$\\
Source width, $w_{s}$&$7.0000\times 10^{-3}$&$6.4\times10^{-6}$\\
Source thickness, $t_{s}$&$3.048\times 10^{-4}$&$6.4\times10^{-6}$\\
Touch gap, $g_{sd}$&$1.080\times 10^{-4}$&$6.4\times10^{-6}$\\
Source amplitude, $dz_{s}$&$1.87\times 10^{-5}$&$3.2\times10^{-6}$\\
\end{tabular}
\end{ruledtabular}
\end{table}

All parameters in Tables~\ref{tab:systematics}
and~\ref{tab:systematics2} are listed with their mean value and error.  As the
information about most of the {\boldmath $\nu$} is limited, a uniform
p.d.f. centered about the mean with a width equal to twice the error is 
assumed for each of the priors $\pi(\mbox{\boldmath$\nu$})$. 

The quantities $G$, $k_{B}$, $\rho_{d}$, and $\rho_{s}$ are taken from
tabulated values~\cite{hagiwara2}.  To obtain the thermal noise 
$\sqrt{\overline{\left|V^{T}\right|^2}}$, the standard deviation of the
off--resonance data (that is, the
noise due to the JFET amplifier) is subtracted in quadrature from the 
standard deviation of the on--resonance data.  The error on
$\sqrt{\overline{\left|V^{T}\right|^2}}$ is estimated in the same way
as the standard deviation of the mean $\sigma/\sqrt{N}$ of the
projected data sets.

Several measurements of the detector mechanical quality factor $Q$ were
made immediately before and after the data sessions,
with the test masses in their fiducial positions.  The
values in Table~\ref{tab:systematics} represent the resulting average 
and standard deviation of these measurements.

The quantity $\sqrt{\int
d^{3}\vec{r}^{\prime}\left|\vec{z}^{F}(\vec{r}^{\prime})\right|^2}$ in
Eq.~\ref{eq:measuredvolt}, the root mean square free modeshape of the
detector integrated over its volume, is computed numerically from a
complete finite element model of the detector mass and is treated as a single
systematic with a uniform error of 10\%. The free modeshape is
normalized to a maximum amplitude of $1~\mu m$.

For a particular set of the {\boldmath $\nu$} in
Table~\ref{tab:systematics2}, a value of the expression $\int
d^{3}\vec{r}^{\prime}\vec{z}^{F}(\vec{r}^{\prime})\cdot
\vec{f}(\vec{r}^{\prime})$ in Eq.~\ref{eq:measuredvolt} is 
obtained in a separate Monte Carlo integration.  In this calculation,
the detector curvature, source mass curvature and source mass
modeshape are described by second order polynomials obtained from fits 
to the survey data.  The detector modeshape is described by a similar
function obtained from the finite element model.  The curvature and modeshape
functions and the {\boldmath $\nu$} are used to calculate bounding 
boxes for the source mass and the small rectangle of the detector
mass.  For a particular position of the source mass, $1 \times 10^{5}$ 
point pairs are thrown.  For those pairs landing within the bounding boxes
a force density $\vec{f}(\vec{r}^{\prime})$ is calculated and 
added to a running total.  A total of 30 different phase positions 
of the source mass between 0 and 2$\pi$ radians are sampled, with 
the range defined by the amplitude $dz_{s}$.

The vertical separation of the test masses is computed
from the touch gap, the test mass thicknesses ($t_{d}, t_{s}$),
and a correction calculated from the functions describing the
test mass surface curvature.  The touch gap ($g_{sd}$) is the 
vertical difference in test mass positions between
where the opposing test mass surfaces touch and where they are in their
fiducial positions.  The error on these quantities 
is taken to be twice the resolution of the translation stages, 
or 6.4~$\mu$m, as two measurements are needed to determine each.  This
is also the case for the errors in the test mass widths ($w_{d},
w_{s}$) and detector length ($l_{d}$).  An exception is the source mass 
tip amplitude ($dz_{s}$), which is the average of five measurements.  
Also, the
accuracy with which the tip of the source mass is aligned with the
back edge of the small detector rectangle is 320 microns, as it is 
aligned by eye and taken to have an error
corresponding to a complete turn of the translation stage screw.  The
accuracy with which the long edge of the source mass is aligned to the
torsion axis is the usual 6.4 microns, as this position is determined
from touch--offs.  All geometry measurements are
repeatable over several weeks to within the resolution of the
translation stages.

\begin{figure}
\includegraphics[width=10cm]{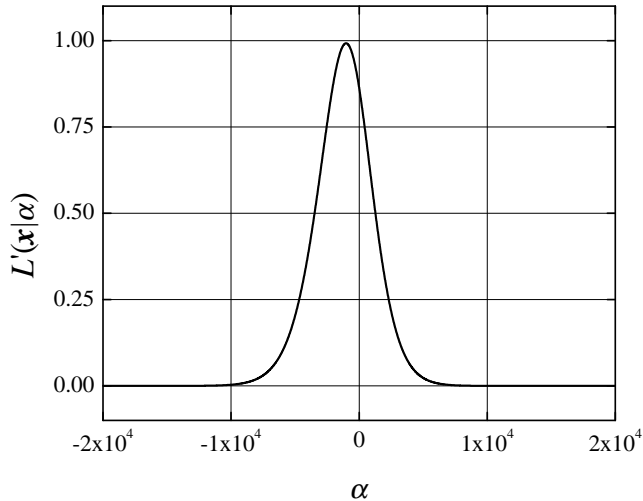}
\caption{\label{fig:likelihood} Likelihood function
  (Eq.~\ref{eq:likelihood}) dependence on $\alpha$ for the 
case $\lambda = 20~\mu$m.}
\end{figure}

Fig.~\ref{fig:likelihood} shows the likelihood function
(Eq.~\ref{eq:likelihood}) as a function of $\alpha$ for the case
$\lambda = 20~\mu$m, a range at which the experiment has good
sensitivity relative to previous experiments.
The factor
$\left(\sigma \sqrt{2\pi e}\right)^{-N}$ has been suppressed
as it has no bearing on the posterior p.d.f.~for $\alpha$ 
(Eq.~\ref{eq:pdf}).  The maximum is shifted slightly toward negative
(repulsive) values of $\alpha$, 
in accordance with the offset of the  mean of the on--resonance data 
with respect to the off--resonance data (Fig.~\ref{fig:summary}).
\subsection{Limits on Yukawa Parameters}
The function in Fig.~\ref{fig:likelihood} is substituted into 
Eq.~\ref{eq:pdf} and integrated numerically (Eq.~\ref{eq:CL}) over
an interval $[\alpha_{\mbox{\scriptsize
    lo}},\alpha_{\mbox{\scriptsize up}}]$ to obtain
the limit on $\alpha$.  The interval is centered around the maximum of 
the likelihood function and the length adjusted by trial and error to 
obtain a confidence level of 95\%.  In Eq.~\ref{eq:pdf}, a 
uniform prior distribution for $\alpha$ is used. For the case 
$\lambda = 20~\mu$m, the resulting 95\% confidence
interval is [$-5.60 \times 10^{3} < \alpha < 3.56 \times 10^{3}$~].  

\begin{figure}[h]
\includegraphics[width = 15cm]{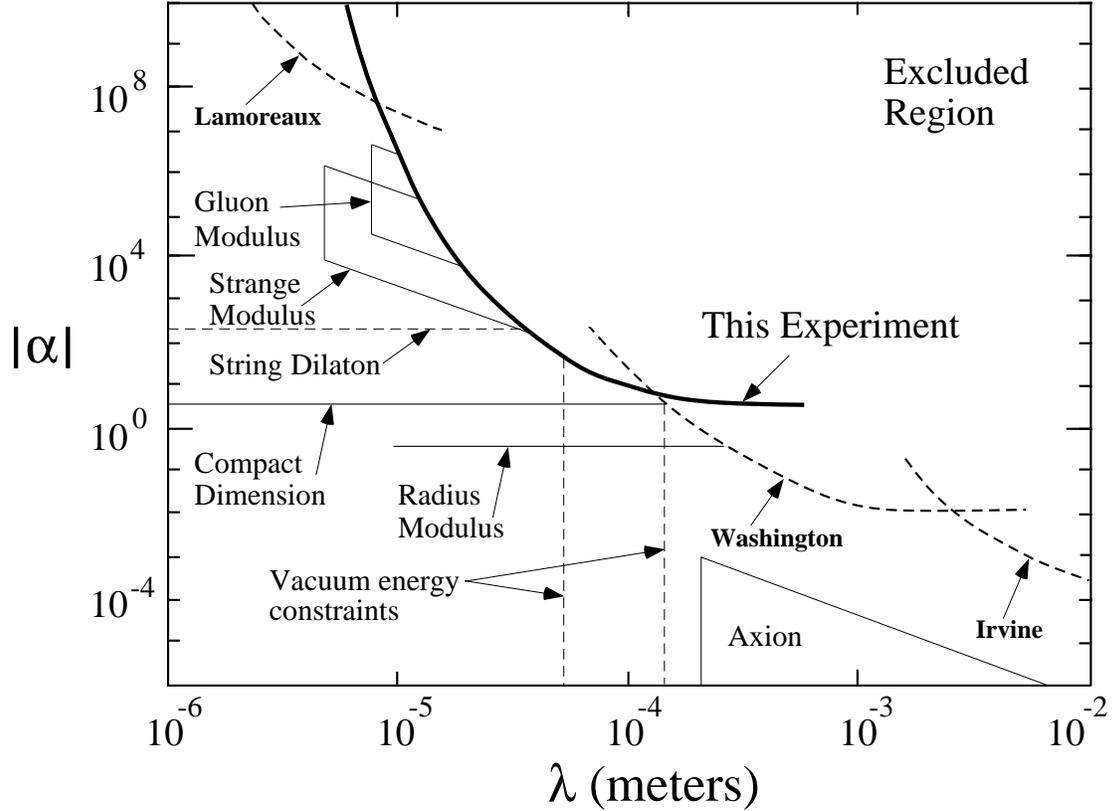}
\caption{\label{fig:limit} 95\% confidence level limit on the Yukawa
  strength $\alpha$ as a function of the range $\lambda$ for this
  experiment (bold solid curve), together with limits from previous
  experiments (bold dashed curves) and theoretical predictions (fine lines).}
\end{figure}

The likelihood function is recomputed and the integration repeated 
for several values of $\lambda$ between 5~$\mu$m and 500~$\mu$m.  The
resulting 95\% confidence level limit curve in the ($\alpha, \lambda$) 
parameter space is shown in Fig.~\ref{fig:limit}.  For each value of
$\lambda$, the slightly weaker limit corresponding to a repulsive
interaction is shown.  The new limit is close
to three orders of magnitude more sensitive than the
previous experimental limits in the range between 10 and 100~$\mu$m. 
\section{Conclusion}
An experimental search for new macroscopic forces at short distances
has been conducted using 1~kHz mechanical oscillators as test 
masses with a stiff conducting shield between them.  No evidence 
for a resonant signal above detector thermal noise has been 
observed.  Based on these measurements, roughly three 
new square decades in the ($\alpha, \lambda$) parameter space
are ruled out in the range 
$10~\mu\mbox{m} < \lambda < 100~\mu\mbox{m}$.  About half of the
remaining parameter space for the gluon and strange modulus forces is
now excluded.  Our limit crosses the predicted line for the dilaton at 
36 microns, corresponding to a lower limit on the dilaton mass of 
$5.5 \times 10^{-3}$~eV.

From Fig.~\ref{fig:limit}, the sensitivity of the experiment
maximizes at about 4 times gravitational strength above 
$\lambda = 200~\mu$m.  For
the planar test mass geometry of the current experiment, the Newtonian signal is roughly an order of magnitude below the thermal
noise at this range.  Detection of this signal should be possible with
more statistics and an optimized geometry. 
The small test masses in this experiment allow for relatively easy
control of the minimum separation, currently limited to
100~$\mu$m by the thickness of the electrostatic shield.  Preliminary tests with
stretched metal membranes suggest that shields of sufficiently low 
compliance can be
made with thicknesses as small as 10~$\mu$m.  Work is underway toward an
experiment with a stretched membrane shield in place of the 
sapphire plate.  If the backgrounds can be controlled, this
experiment could improve limits between 10 and 50~$\mu$m by at least another order of magnitude.
\begin{acknowledgments}
The authors would like to thank Elizabeth Lagae for her continuing 
work in the laboratory, and Cole Briggs, Tracy Buxkemper, Leslie
Czaia, Hans Green, Sid Gustafson, and Hans Rohner of the University of 
Colorado and JILA instrument shops for technical assistance.  
This work is supported by NSF grant PHY00-71029.
\end{acknowledgments}
\appendix
\section{Calibration}
The mean signal from the lock-in amplifier data can be converted to
observed force by direct comparison to the detector thermal noise.
The mean square thermal displacement of the detector mass
$\overline{\left|\vec{z}^{T}(\vec{r})\right|^2}$ at the position
$\vec{r}$ of the capacitive transducer probe is related to the temperature
$T$ via the equipartition theorem:
\begin{equation}
\frac{1}{2}m(\vec{r})\overline{\left|\vec{z}^{T}(\vec{r})\right|^2}
\omega_{0}^{2} = \frac{1}{2} k_{B}T.
\label{eq:equipartition}
\end{equation} 
Here, $\omega_{0}$ is the resonant frequency of the 5th detector mode and
$m(\vec{r})$ is the modal mass of the detector oscillator at
the position of the probe.

For a distributed mass with free modal displacement
$\vec{z}^{F}(\vec{r}^{\prime})$ at an arbitrary point $\vec{r}^{\prime}$
driven on resonance, the driven 
displacement amplitude $d\vec{z}^{D}(\vec{r}^{\prime})$ due to a
differential force amplitude $d\vec{F}(\vec{r}^{\prime})$
is given by
\begin{equation}
d\vec{z}^{D}(\vec{r}^{\prime}) =
-j\frac{Q}{\omega_{0}^{2}m(\vec{r}^{\prime})}\vec{z}^{F}(\vec{r}^{\prime})\frac{\vec{z}^{F}(\vec{r}^{\prime})\cdot
  d\vec{F}(\vec{r}^{\prime})}{\left|\vec{z}^{F}(\vec{r}^{\prime})\right|^{2}},
\label{eq:arbdriveninstdisp}
\end{equation} 
where $Q$ is the detector mechanical quality factor and the dot
product picks out the component of the force parallel to the modal 
displacement. (Note the use of $e^{+j\omega t}$ time dependence.)
The differential amplitude at the probe is
\begin{equation}
d\vec{z}^{D}(\vec{r}) = \left|d\vec{z}^{D}(\vec{r}^{\prime})\right|\frac{\vec{z}^{F}(\vec{r})}{\left|\vec{z}^{F}(\vec{r}^{\prime})\right|}.
\label{eq:transdriveninstdisp}
\end{equation}
Substituting Eq.~\ref{eq:transdriveninstdisp} into 
Eq.~\ref{eq:arbdriveninstdisp} and integrating over
$\vec{r}^{\prime}$ yields the amplitude of the detector at the
position of the transducer due the total force on the
detector:
\begin{equation}
\vec{z}^{D}(\vec{r}) =
-j\frac{Q}{\omega_{0}^{2}\rho_{d}}\vec{z}^{F}(\vec{r})\frac{\int d^{3}\vec{r}^{\prime}\vec{z}^{F}(\vec{r}^{\prime})\cdot
  \vec{f}(\vec{r}^{\prime})}{\int d^{3}\vec{r}^{\prime}\left|\vec{z}^{F}(\vec{r}^{\prime})\right|^2}.
\label{eq:transdrivendisp}
\end{equation}
Here the force amplitude density $\vec{f}(\vec{r}^{\prime})$ 
is defined by $\vec{f}(\vec{r}^{\prime})\equiv
d\vec{F}(\vec{r}^{\prime})/d^{3}\vec{r}^{\prime}$, $\rho_{d}$ is the
detector mass density, and the definition of the modal mass
\begin{equation}
m(\vec{r}) = \frac{\rho_{d} \int
  \left|\vec{z}^{F}(\vec{r}^{\prime})\right|^2 d^{3}\vec{r}^{\prime}}{\left|\vec{z}^{F}(\vec{r})\right|^2}
\label{eq:modalmass}
\end{equation}
has been used.

Combining Eqs.~\ref{eq:equipartition},~\ref{eq:transdrivendisp},
and~\ref{eq:modalmass} yields
\begin{equation}
\frac{\vec{z}^{D}(\vec{r})}{\sqrt{\overline{\left|\vec{z}^{T}(\vec{r})\right|^2}}}=
  \frac{Q}{j\omega_{0}\sqrt{k_{B}T\rho_{d}}}\frac{\vec{z}^{F}(\vec{r})}{\left|\vec{z}^{F}(\vec{r})\right|}\frac{\int d^{3}\vec{r}^{\prime}\vec{z}^{F}(\vec{r}^{\prime})\cdot
  \vec{f}(\vec{r}^{\prime})}{\sqrt{\int d^{3}\vec{r}^{\prime}\left|\vec{z}^{F}(\vec{r}^{\prime})\right|^2}}.
\label{eq:measureddisp}
\end{equation}
The linear response of the capacitive transducer insures that the
ratio $\vec{z}^{D}(\vec{r})/\sqrt{\overline{\left|\vec{z}^{T}(\vec{r})\right|^2}}$
is related to the measured voltages on the lock-in amplifier by
\begin{equation}
\frac{\left|\vec{z}^{D}(\vec{r})\right|}{\sqrt{\overline{\left|\vec{z}^{T}(\vec{r})\right|^2}}}=\frac{\overline{V^{D}}}{\sqrt{\overline{\left|V^{T}\right|^2}}},
\label{eq:zvratio}
\end{equation}
where $\overline{V^{D}}$ is the mean of the distribution of voltages
in the data sample and $\sqrt{\overline{\left|V^{T}\right|^2}}$ is the 
component of the standard deviation of the same data sample due to 
detector thermal noise.  Both quantities are evaluated at the
phase of interest.  Substitution of Eq.~\ref{eq:zvratio} into
Eq.~\ref{eq:measureddisp} yields Eq.~\ref{eq:measuredvolt}.
\section{Likelihood Function}
In addition to the unknown parameter $\alpha$ corresponding to the
Yukawa strength, the signal depends on a set of 
systematics {\boldmath $\nu$} which includes test mass geometry, density, 
detector temperature and mechanical properties.  The effects of 
uncertainties in these quantities can be accounted 
for in the posterior p.d.f. (Eq.~\ref{eq:pdf}) by integrating 
over the systematics:
\begin{equation}
p(\alpha|\mbox{\boldmath$x$}) = \int p(\alpha,\mbox{\boldmath$\nu$}|\mbox{\boldmath$x$})d\mbox{\boldmath$\nu$}.
\end{equation}
Assuming the prior joint p.d.f. for $\alpha$ and {\boldmath$\nu$} factorizes,
this is equivalent to replacing the likelihood function $L(\mbox{\boldmath$x$}|\alpha)$
with 
\begin{equation}
L^{\prime}(\mbox{\boldmath$x$}|\alpha) = \int L(\mbox{\boldmath$x$}|\alpha,\mbox{\boldmath$\nu$})\pi(\mbox{\boldmath$\nu$})d\mbox{\boldmath$\nu$}.
\label{eq:likelihoodA}
\end{equation}
Here, $\pi(\mbox{\boldmath$\nu$})$ is the prior p.d.f. of
the {\boldmath$\nu$}. The function
$L(\mbox{\boldmath$x$}|\alpha,\mbox{\boldmath$\nu$})$ is the joint
p.d.f. for the data, regarded as a function of both the unknown Yukawa
strength $\alpha$ and the systematics {\boldmath $\nu$}:
\begin{equation}
L(\mbox{\boldmath$x$}|\alpha,\mbox{\boldmath$\nu$}) = \prod_{i}F(x_{i}|\alpha,\mbox{\boldmath$\nu$}).
\label{eq:integrand}
\end{equation}

For the present experiment, the data $x_{i}$ are the
sampled voltages from the lock-in amplifier, which are 
Gaussian distributed (Fig.~\ref{fig:raw}).  Therefore the 
function $F$ is taken to be the
Gaussian distribution
$G(x_{i}|\mu(\alpha,\mbox{\boldmath$\nu$}),\sigma)$, where
$\mu(\alpha,\mbox{\boldmath$\nu$})$ is the {\it predicted} mean
voltage for a given $\alpha$ and set of systematics {\boldmath $\nu$},
and $\sigma$ is the observed standard deviation:
\begin{equation}
F(x_{i}|\alpha,\mbox{\boldmath$\nu$}) = \left(\frac{1}{\sigma
    \sqrt{2\pi}}\right)\exp
    \left[-(x_{i}-\mu(\alpha,\mbox{\boldmath$\nu$}))^2/2\sigma^{2} \right].
\end{equation}
The product is
\begin{eqnarray}
\prod_{i}F(x_{i}|\alpha,\mbox{\boldmath$\nu$}) & = &
\left(\frac{1}{\sigma
    \sqrt{2\pi}}\right)^{N}\exp
\left[-\frac{1}{2\sigma^{2}}\sum_{i=1}^{N}(x_{i}-\mu(\alpha,\mbox{\boldmath$\nu$}))^2
  \right]\nonumber \\
 & = & \left(\frac{1}{\sigma
    \sqrt{2\pi e}} \right)^{N} \exp\left[-\frac{(\bar{x}-\mu(\alpha,\mbox{\boldmath$\nu$}))^2}{2(\sigma/\sqrt{N})^{2}}\right],
\label{eq:prodA}
\end{eqnarray}
where $\bar{x}$ is the average of the voltages $x_{i}$ and $N$ is the effective
number of uncorrelated samples (the value of $N$ such that
$\sigma/\sqrt{N}$ is the standard deviation of the mean of the actual data).  

The predicted mean $\mu(\alpha,\mbox{\boldmath$\nu$})$ can be replaced
by Eq.~\ref{eq:measuredvolt} for $\overline{V^{D}}$ if 
$\vec{f}(\vec{r}^{\prime})$
is identified with the {\it theoretical} value of the Yukawa force
density for a test mass geometry corresponding to a particular set of 
parameters {\boldmath$\nu$}.  As a Yukawa force density, 
$\vec{f}(\vec{r}^{\prime})$ is linear in the
parameter $\alpha$, the gravitational constant $G$, and the source mass
and detector mass densities $\rho_{s}$ and $\rho_{d}$:
$\vec{f}(\vec{r}^{\prime}) = \alpha G \rho_{s} \rho_{d}
\vec{\mbox{f}}(\vec{r}^{\prime})$.  Substituting into
Eq.~\ref{eq:measuredvolt} and factoring these terms out of the
integral yields
\begin{equation}
\mu(\alpha,\mbox{\boldmath$\nu$}) = \left(\alpha G \rho_{s} \sqrt{\rho_{d}}\right)
  \sqrt{\overline{\left|V^{T}\right|^2}}
  \left(\frac{Q}{\omega_{0}\sqrt{k_{B}T}}\right) \frac{\int d^{3}\vec{r}^{\prime}\vec{z}^{F}(\vec{r}^{\prime})\cdot
  \vec{\mbox{f}}(\vec{r}^{\prime})}{\sqrt{\int d^{3}\vec{r}^{\prime}\left|\vec{z}^{F}(\vec{r}^{\prime})\right|^2}}.
\label{eq:model2}
\end{equation}
Substituting Eq.~\ref{eq:model2} into Eq.~\ref{eq:prodA} and
the result into Eq.~\ref{eq:likelihoodA} yields the likelihood
function $L^{\prime}(\mbox{\boldmath$x$}|\alpha)$.


\begin{thebibliography}{99}
\bibitem{mostepanenko} Another estimate (Ref.~\cite{bordag} above) which 
  considers corrections to the Casimir force not considered by the
  authors, such as those for surface roughness and finite temperature, 
  suggests this limit to be about an order of magnitude more
  sensitive.
\end{thebibliography}
\end{document}